\documentclass[fleqn]{article}
\usepackage[margin=1.5in]{geometry} %
\usepackage{mathtools}
\usepackage{amssymb}
\usepackage{mathrsfs}
\usepackage{bbm}
\usepackage{cite}%
\usepackage{graphicx}
\usepackage{xcolor}
\usepackage{nicefrac}
\usepackage[hidelinks]{hyperref}
\usepackage{cleveref}
\newcommand{\rep}[1]{\ensuremath{\boldsymbol{#1}}}

\DeclareMathOperator{\re}{Re}
\DeclareMathOperator{\im}{Im}

\DeclareMathOperator{\diag}{diag}

\newcommand{\D}{\mathrm{d}}
\newcommand{\I}{\mathrm{i}}

\newcommand{\SU}[1]{\ensuremath{\mathrm{SU}(#1)}}
\newcommand{\SL}[1]{\ensuremath{\mathrm{SL}(#1)}}
\newcommand{\U}[1]{\ensuremath{\mathrm{U}(#1)}}
\newcommand{\Z}[1]{\ensuremath{\mathbbm{Z}_{#1}}} 
\newcommand{\Tprime}{\ensuremath{\text{T}'}}
\def\mytitle{A note on the predictions of models with modular flavor symmetries}
\title{\mytitle}
\begin{document}
\begin{titlepage}
\vspace*{-3.0cm}

\begin{flushright}
UCI--TR--2019--22%
\\
TUM--HEP 1223/19
\end{flushright}

\vspace*{1.0cm}

\begin{center}
{\Large\bfseries\mytitle}

\vspace{1cm}

\textbf{%
Mu--Chun Chen$^{a}$, Sa\'ul Ramos--S\'anchez$^{b,c}$, Michael Ratz$^{a}$
}
\\[8mm]
\textit{$^a$\small
~Department of Physics and Astronomy, University of California, Irvine, CA 92697-4575 USA
}
\\[5mm]
\textit{$^b$\small Instituto de F\'isica, Universidad Nacional Aut\'onoma de M\'exico, POB 20-364, Cd.Mx. 01000, M\'exico}\\
\textit{$^c$\small Physik Department T75, Technische Universit\"at M\"unchen, James-Franck-Stra\ss e 1, 85748 Garching, Germany}
\end{center}

\vspace*{1cm}

\begin{abstract}
Models with modular flavor symmetries have been thought to be highly predictive.
We point out that these predictions are subject to corrections from
non--holomorphic terms in the Lagrangean. Specifically, in the models discussed
in the literature, the K\"ahler potential is not fixed by the symmetries, for
instance. The most general K\"ahler potential consistent with the symmetries of
the model contains additional terms with additional parameters, which reduce the
predictive power of these constructions. We also comment on how one may
conceivably retain the predictivity.
\end{abstract}
\vspace*{1cm}
\end{titlepage}

\section{Introduction}

Recently a rather exciting observation has been
made~\cite{Feruglio:2017spp,Criado:2018thu}: nine neutrino parameters can be
predicted from only three input parameters. The crucial ingredients of the
corresponding model are modular flavor symmetries. The point of this paper is to
show that these models actually have additional parameters which have not been
taken into account in the models in the recent literature. We also comment on
possible ways to retain control over these parameters.

To understand the main point of our paper, recall that the predictions of
these models come from the fact that the \emph{superpotential} is fixed by the
modular transformations. However, the superpotential only contains the physical
parameters if the fields appearing there are ``physical'', i.e.\ canonically
normalized. As we shall see, the K\"ahler potential, which contains the
information about the fields, is not at all fixed by the symmetries and
transformation properties of the models. This is why the modular transformations
alone do not allow one to make such remarkable predictions, as we shall discuss
in more detail in what follows.

\section{Modular flavor symmetries}

Modular flavor symmetries have so far
only been discussed in the supersymmetric context. There, they are modular
transformations which act on a so--called modulus $\tau$ and ``matter''
superfields $\varphi^{(j)}$ according to
\cite{Feruglio:2017spp}
\begin{subequations}\label{eq:ModularTrafo}
\begin{align}
 \tau&~\mapsto~\frac{a\,\tau+b}{c\,\tau+d}~=:~\gamma\,\tau\;,\label{eq:ModularTrafoTau}\\
 \phi^{(j)}
 &~\mapsto~(c\,\tau+d)^{-k_j}\,\rho^{(j)}(\gamma)\,\phi^{(j)}\;,
 \label{eq:ModularTrafoMatter}
\end{align}
\end{subequations}
where $a$, $b$, $c$ and $d$ are the $\Gamma \equiv \SL{2,\mathbbm{Z}}$
parameters satisfying, by definition, $a\,d-b\,c=1$ and $\rho^{(j)}$ is the
representation matrix of some quotient group $\Gamma_N=\Gamma/\Gamma(N)$. $-k_j$
denotes the so--called modular weight. The collection of chiral superfields will
be denoted $\Phi=(\tau,\phi^{(1)},\ldots,\phi^{(F)})$.

The modular group $\overline{\Gamma}=\Gamma/\Z2$ is generated by
\begin{equation}
 S~=~\begin{pmatrix} 0 & 1 \\ -1 &0 \end{pmatrix}
 \quad\text{and}\quad
 T~=~\begin{pmatrix} 1 & 1 \\ 0 &1 \end{pmatrix}\;,
\end{equation}
which correspond to the transformations
\begin{equation}
 \tau~\xmapsto{~S~}~-\frac{1}{\tau}
 \quad\text{and}\quad
 \tau~\xmapsto{~T~}~\tau+1\;.
\end{equation}
These generators satisfy
\begin{equation}
 S^2~=~(S\,T)^3~=~\mathbbm{1}\;.
\end{equation}

It is straightforward to verify that
\begin{align}
 \left(-\I\,\tau+\I\,\bar\tau\right)^{-k}&~\xmapsto{(\ref{eq:ModularTrafo})}~
 \left(\left(c\,\tau+d\right)\,\left(c\,\bar\tau+d\right)\right)^{k}\,
 \left(-\I\,\tau+\I\,\bar\tau\right)^{-k}\;.
\end{align}
Therefore, the combination
\begin{align}
 \left(-\I\,\tau+\I\,\bar\tau\right)^{-k_j}\,
 	\left(\phi^{(j)*}\phi^{(j)}\right)_{\rep{1}}
\label{eq:simpleK}
\end{align}
is invariant under modular transformations. Here, the notation
$\left(\cdots\right)_{\rep{1}}$ indicates a contraction to a $\Gamma_N$
\rep{1}--plet, i.e.\ to an invariant under $\Gamma_N$. 
However, as we shall see below, this is not the only invariant. 

\section{Additional parameters from non--holomorphic terms}

The fact that there are additional terms in the K\"ahler potential has been
already noted in \cite{Feruglio:2017spp}. The existence of additional terms
already follows from the observation that the predicted parameters run. Running
of couplings in supersymmetric theories can be understood as corrections to the
K\"ahler potential. On the other hand, the superpotential is protected by
holomorphicity, which is reflected by the non--renormalization theorems. As we
shall see, the most general K\"ahler potential consistent with the symmetries
has numerous additional parameters.

We will base our discussion on Model 1 of \cite{Feruglio:2017spp}, which
has the finite quotient symmetry $\Gamma_3 \simeq A_4$.  However, the
analogous statements apply to the follow--up models in the literature such as 
\cite{Criado:2018thu,Kobayashi:2018scp,Novichkov:2018ovf,Kobayashi:2018bff,deAnda:2018ecu,Kobayashi:2018wkl,Ding:2019xna,Kobayashi:2019mna}.
The Higgs and lepton sector of the model is specified in
\cref{tab:FeruglioModel}. 

\begin{table}[htb]
\centering
$\begin{array}{|*{6}{c|}|c|}
\hline
  & (E_1^c,E_2^c,E_3^c) &  N & L & H_d & H_u & \varphi\\
\hline 
 \SU{2}_\mathrm{L}\times\U1_Y & \rep{1}_1 & \rep{1}_0 & \rep{2}_{-\nicefrac{1}{2}}
  & \rep{2}_{-\nicefrac{1}{2}} & \rep{2}_{\nicefrac{1}{2}} & \rep{1}_0\\
\hline
 \Gamma_3 & (\rep{1},\rep{1'},\rep{1''}) & \rep{3} & \rep{3} & \rep{1} & \rep{1}
 & \rep{3} \\
\hline
 k & (k_{E_1},k_{E_2},k_{E_3}) & k_N & k_L & k_{H_d} & k_{H_u} & k_\varphi \\
\hline
\end{array}$
\caption{Model 1 of \cite{Feruglio:2017spp}. $E_i^c$, $L$, $H_u$ and $H_d$ are
the superfields of the charged leptons, left--handed douplets, up--type Higgs and
down--type Higgs, respectively.}
\label{tab:FeruglioModel}
\end{table}

As the author of \cite{Feruglio:2017spp} has pointed out, the charged fermion masses are obtained by
adjusting three parameters. The nontrivial predictions of this model are on the
neutrino parameters, which come form the Weinberg operator
\begin{equation}
 \mathscr{W}_\nu~=~\frac{1}{\Lambda}\left[\left(H_u\cdot L\right)\,Y\,
 \left(H_u\cdot L\right)\right]_{\rep{1}}\;.
\end{equation}
Here, $Y$ is a triplet of modular functions of weight 2,
$Y=\left(Y_1,Y_2,Y_3\right)^T$. The K\"ahler potential of the charged leptons is
taken to be 
\begin{equation}\label{eq:NaiveKahlerL}
 K_L~=~\left(-\I\,\tau+\I\,\bar\tau\right)^{-1}\,
 L^\dagger\,L\;.
\end{equation}
Here the modular weights of the leptons are $-1$ (corresponding to $k_L=1$) and $H_u$ has 
zero weight ($k_{H_u}=0$). The neutrino mass matrix is then given by
\begin{equation}\label{eq:mnu_modular}
 m_\nu~=~\frac{v_u^2}{\Lambda}\begin{pmatrix}
  2Y_1(\tau) & -Y_3(\tau) & -Y_2(\tau) \\
  -Y_3(\tau) & 2Y_2(\tau) & -Y_1(\tau) \\
  -Y_2(\tau) & -Y_1(\tau) & 2Y_3(\tau) 
 \end{pmatrix}\;.
\end{equation}
The crucial point is that this matrix has only three free real parameters:
$\Lambda$, $\re\tau$ and $\im\tau$. On the other hand, the charged lepton Yukawa
coupling is diagonal in this model. Therefore, the mass matrix
\eqref{eq:mnu_modular} fixes nine observables: the three neutrino mass
eigenvalues, three mixing angles, the so--called Dirac $\mathcal{CP}$ phase
and two Majorana phases. In \cite{Feruglio:2017spp,Criado:2018thu} values of
$\tau$ that gives rise to realistic neutrino masses and mixing 
angles are specified. This is a spectacular result. Three real input parameters, $\Lambda$,
$\re\tau$ and $\im\tau$, pin down three mass eigenvalues, three mixing angles
and three phases. That is, this setting appears to make six nontrivial
predictions, which agree amazingly well with observation (so far).

In more detail, the MNS matrix is the mismatch of the unitary
transformations that diagonalize the neutrino mass matrix and the charged
lepton Yukawa coupling matrix, respectively,
\begin{align}\label{eq:0thOrderDiagonalization}
 U_\nu^T\,m_\nu\,U_\nu~=~\diag(m_1,m_2,m_3)\quad\text{and}\quad
U^\dagger_{e_\mathrm{L}}\,Y_e\,Y_e^\dagger\, U_{e_\mathrm{L}}~=~\diag(y_e^2,y_\mu^2,y_\tau^2)\;.
\end{align}
That is, $U_\mathrm{MNS}^{(0)}=U_{e_\mathrm{L}} \, U_\nu^T$, and since in the original
Lagrangean $Y_e$ is diagonal, $U_\mathrm{MNS}^{(0)}=U_\nu^T$. The first term
depends on nine physical parameters,
\begin{align}
 m_\nu~=~U_\nu^*
 \,\diag(m_1,m_2,m_3)\,U_\nu^\dagger
 \;,\quad\text{where}~
 U_\nu=U_\nu(\theta_{12},\theta_{13},\theta_{23},\delta,\varphi_1,\varphi_2,
 \dots)
\end{align}
with $\theta_{ij}$ denoting the three mixing angles, $\delta$ the Dirac
$\mathcal{CP}$ phase, $\varphi_i$ the two Majorana phases, and the omission
``$\dots$'' stands for three unphysical phases.

This parameter counting assumes that the K\"ahler potential is given by
\eqref{eq:NaiveKahlerL}. However, the modular symmetries do not fix the form of the K\"ahler potential. 
Rather, the full K\"ahler potential includes additional terms beyond the one given in \eqref{eq:NaiveKahlerL},
\begin{align}
 K&~=~
 \alpha_0\,\left(-\I\,\tau+\I\,\bar\tau\right)^{-1}\,\left(\overline{L}\,L\right)_{\rep{1}}
 +\sum_{k=1}^7\alpha_k
 \left(-\I\,\tau+\I\,\bar\tau\right)\,\left(Y\,L\,\overline{Y}\,\overline{L}
 \right)_{\rep{1}, \, k}+\dots  \label{eq:DeltaK} \; .
\end{align}
Here we have summed over all singlet contractions (specified by subscript $k$),
and $\alpha_0$ can be absorbed in a redefinition of the fields. Some of the
relevant contractions are given by
$\left\{\left(\overline{Y}\,\overline{L}\right)_{\rep{3}^{(j)}}^T
\left(Y\,L\right)_{\rep{3}^{(i)}} \right\}_{i,j\in\{1,2\}}$ and the invariant
contractions of the one--dimensional contractions
$\left(Y\,L\right)_{\rep1,\rep1',\rep1''}$ with appropriate
conjugates.\footnote{Notice that the conjugate of $\left(Y\,L\right)_{\rep1'}$,
$\left[\left(Y\,L\right)_{\rep1'}\right]^*$, transforms as $\rep{1}''$.}
Specifically, the first three terms in the expansion \eqref{eq:DeltaK} are 
\begin{align}
 \Delta K ~=~&\alpha_1\,
 \left(\overline{Y}\,\overline{L}\right)_{\rep{3}^{(1)}}^T
 \left(Y\,L\right)_{\rep{3}^{(1)}} 
 +\alpha_2\,
 \left(\overline{Y}\,\overline{L}\right)_{\rep{3}^{(2)}}^T
 \left(Y\,L\right)_{\rep{3}^{(2)}} \notag\\
 &{}+\alpha_3\,\left[
 \left(\overline{Y}\,\overline{L}\right)_{\rep{3}^{(1)}}^T
 \left(Y\,L\right)_{\rep{3}^{(2)}} 
 +\left(\overline{Y}\,\overline{L}\right)_{\rep{3}^{(2)}}^T
 \left(Y\,L\right)_{\rep{3}^{(1)}} \right]+\ldots\;.\label{eq:DeltaK1-3}
\end{align}
Note that all the terms are on the same footing, there is a priori no reason
why, say, the $\alpha_0$ term should be referred to as the leading term and the
others as ``corrections''.

Once we add the other fields of the model, even more terms will have to be added. For
instance, the above model \cite{Feruglio:2017spp,Criado:2018thu} also introduces
a flavon $\varphi$ (cf.\ \Cref{tab:FeruglioModel}). Therefore, we can add
further terms to the K\"ahler potential of the form
\begin{align}\label{eq:DeltaK2}
 \Delta K&~=~\sum_i\beta_i
 \left(-\I\,\tau+\I\,\bar\tau\right)^{-k_L-k_\varphi}\,
 \left(\varphi\,L\,\overline{\varphi}\,\overline{L}
 \right)_{\rep{1}, \, i}\;,
\end{align}
where we sum over all $A_4$--invariant contractions.

The impacts of these additional terms can be significant. Suppose one has
derived predictions on the neutrino parameters based on the K\"ahler potential
\eqref{eq:NaiveKahlerL}. The additional terms will modify the K\"ahler metric,
\begin{equation}
 K^{i\bar\jmath}_L~=~\frac{\partial^2K}{\partial L_i\,\partial\overline{L}_{\bar\jmath}}\;. 
\end{equation}
This metric has to be diagonalized, 
\begin{equation}
 K_L~=~U_L^\dagger\,D^2\,U_L\;,
\end{equation}
where $U_L$ is unitary and $D$ is diagonal and positive. Therefore, the
canonically normalized fields are 
\begin{align}
 \widehat{L}~=~D\,U_L\,L\quad \text{or equivalently}\quad
 L~=~U_L^\dagger\,D^{-1}\,\widehat{L}\;.
\end{align}

\begin{figure}[!htb]
 \centering
 \includegraphics[scale=1.2]{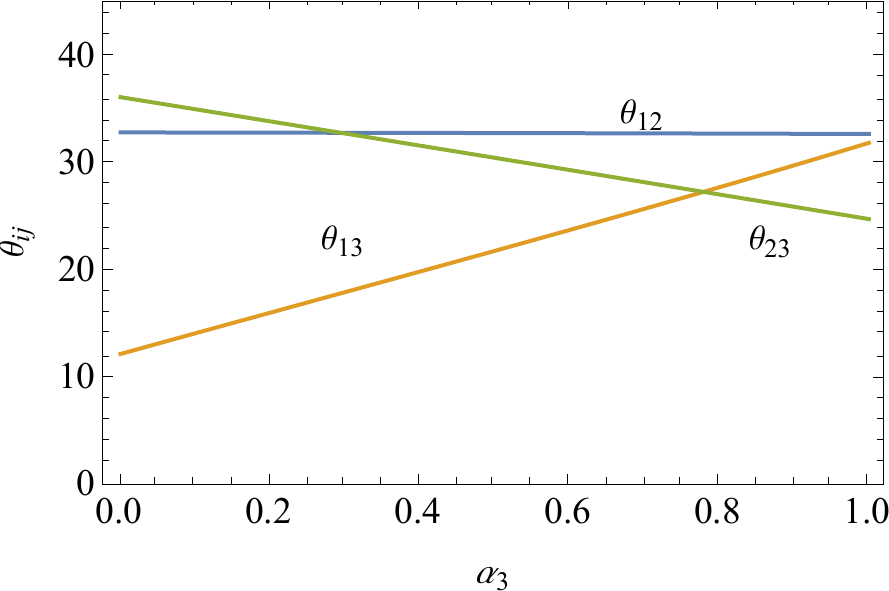}
 \caption{Dependence of the mixing angles on the additional parameter $\alpha_3$
 (cf.\ \Cref{eq:DeltaK1-3}).}
 \label{fig:MFS-KaehlerPlot1}
\end{figure}

After adding the $\alpha_{i>0}$ contributions and transforming the fields back to
canonical normalization, we need to diagonalize
\begin{subequations}
\begin{align}
 \widehat{U}_\nu^T\,D^{-1}\,U_L^*\,m_\nu\,U_L^\dagger\,D^{-1}\,\widehat{U}_\nu~&=~\diag(m_1,m_2,m_3)
 \;,\\
 \widehat{U}_e^\dagger\,D^{-1}\,U_L^*\,Y_e\,Y_e^\dagger\,U_L^T\,D^{-1}\,
 \widehat{U}_e~&=~\diag(y_e^2,y_\mu^2,y_\tau^2)\;.
\end{align}
\end{subequations}
This is to be compared with \eqref{eq:0thOrderDiagonalization}. We see that if
$D$ is proportional to the unit matrix, there would be no effect, i.e.\  the
original mixing matrix $U_{\text{L}}$ would still do the job of diagonalizing
$m_{\nu}$ and thus the predicted values for the neutrino mixing parameters 
based solely on $\alpha_{0}$ contribution remain valid. However, for the
contributions given in \eqref{eq:DeltaK}, $D$ is generically not proportional to
the unit matrix, and consequently the predicted values for the mixing angles get
modified significantly. Our numerical analysis reveals that they are of the
order
\begin{align}\label{eq:DeltaTheta}
 \Delta \theta_{ij}^{(k)}~=~\alpha_k\cdot (1\dots10)^\circ\;,
\end{align} 
and similarly for the $\mathcal{CP}$ phases. This is illustrated in
\Cref{fig:MFS-KaehlerPlot1} for $k=3$.
Analytic formulae that allow one to evaluate the impact of these corrections
have been derived in \cite{Chen:2012ha,Chen:2013aya}.  They confirm our result
as given in \eqref{eq:DeltaTheta}. Importantly, these corrections are in general
much larger than the corrections from RGE running and supersymmetry breaking
which have been worked out in \cite{Criado:2018thu}.  

Altogether we see that in models with modular flavor symmetries the
specification of $\tau$ and $\Lambda$ is \emph{not} sufficient to determine the
neutrino parameters. There exist many additional parameters, and, as a
consequence, the number of free parameters is generically larger than the number
of predictions.

\section{Discussion}

The findings of the previous section should not be surprising. The salient
properties of the models with modular flavor symmetries rely on the
holomorphicity of the superpotential. However, the K\"ahler potential does not
have these properties. Moreover, these symmetries are nonlinearly realized.

How can one conceivably control the K\"ahler potential better? This will be
possible if one derives the modular flavor symmetries from some more complete
setting. As is well known, these symmetries come from tori. Thus one expects
that there will be interpretations of these symmetries in models with extra
dimensions.

Most prominently, modular symmetries appear in string theory. The existence of
some non--Abelian symmetries has been already noted in \cite{Ferrara:1989qb},
and more recently studied in more detail in~\cite{Baur:2019kwi,Baur:2019iai}. In
particular, the \Z3 orbifold, which also has (in the absence of so--called
discrete Wilson lines) a $\Delta(54)$ flavor symmetry~\cite{Kobayashi:2006wq},
has a $\Tprime$ modular flavor symmetry~\cite{Baur:2019iai}. Given these
results, it is tempting to speculate that an $A_4$ modular flavor symmetry could
originate from the $\mathbb{T}^2/\Z2$ orbifold, where the four twisted string
states  form a $(\rep{3}+\rep{1})$ reducible representation. 

Note that in string theory, the notation is usually somewhat different
(cf.\ e.g.~\cite{Ibanez:1992hc}). Instead of denoting the modulus $\tau$ and
demanding that its \emph{imaginary} part transforms as a real scalar and its
\emph{real} part as a pseudoscalar, many string theorists prefer to consider $T$
instead of $\tau = \I\,T$. Then the real part transforms as scalar and has
often the interpretation of volume. The imaginary part is sometimes referred to
as $T$--axion. The transformation of $T$ and the matter fields 
under $\gamma \in \Gamma_N$ then reads  
\begin{subequations}\label{eq:TdualityStrings}
\begin{align}
 T&~\mapsto~\frac{a\,T-\I\,b}{\I\,c\,T+d}\;,\\
 \varphi^{(j)}&~\mapsto~\left(\I\,c\,T+d\right)^{n_j}\,
 \rho^{(j)}(\gamma)\,\varphi^{(j)}\;,
\end{align}
\end{subequations}
where the $n_j=-k_j$ are the modular weights. 

In contrast to the bottom--up models, in many string theory compactifications
the modular weights are not free parameters but  can be computed from  other
data of the models. They are used to derive approximate expressions for the
K\"ahler potential. For example, by considering string scattering amplitudes in
heterotic orbifold compactifications (although this result is more general; see
e.g.~\cite{Conlon:2006tj}) and the so--called large volume limit $\re T \gg 1$, 
it has been found that the leading  contribution to the K\"ahler potential for
the matter fields is given by~\cite{Kaplunovsky:1995jw}
\begin{equation}
\label{eq:stringyK}
 K ~\supset~\sum_\ell F_{i\ell}^*(\overline T) F_{\ell j}(T) \left(T+\overline{T}\right)^{n_j}\, \overline \varphi^{(i)}\varphi^{(j)}\,,
\end{equation}
where the modular weights $n_j$ are derived from the oscillator quantum numbers and
the twist of the fields $\varphi^{(j)}$, and turn out to be (mostly) nonpositive.
$F_{\ell j}$ are arbitrary holomorphic functions, building a non--degenerate
matrix that fix the basis of the field space.  Although these functions are
typically chosen
as $F_{\ell j} = \delta_{\ell j}$  for all $j$ and $\ell$, one
may in principle also consider modular forms of nontrivial modular weight
$n_{F_{\ell j}}$. Modular invariance of the K\"ahler potential would then imply
that $n_j$ must be replaced by  $n_j+n_{F_{\ell j}}$ in \Cref{eq:stringyK}.
If we suppose that $F_{\ell j}=\delta_{\ell j}Y(T)$ for $\varphi^{(j)}=L$, the terms of
the K\"ahler potential~\eqref{eq:DeltaK} with $k\neq 0$ are recovered with no
additional suppression. Note however that the functions $F_{\ell j}$ can be 
absorbed in field redefinitions at the expense of altering the superpotential
couplings.

It is known that the K\"ahler potential~\eqref{eq:stringyK} receives additional
contributions (see e.g.~\cite{Antoniadis:1994hg}). E.g.\ for  string
compactifications where matter arises from bulk fields, the K\"ahler potential
can  be expressed as  $K = -\ln (T + \overline{T} - |\varphi^{(j)}|^2)$,  which
yields~\eqref{eq:simpleK} only in the large volume limit. However,
the best--fit point for phenomenology in the model discussed ($\re T \approx 1$)
violates this limit. It should also be noted that in string compactifications
the superpotential usually transforms nontrivially, and has modular weight $-1$.

Furthermore,  as is well known, string theory is in principle very predictive.
However, in concrete examples it is nontrivial to make precise predictions. This
is because string models leave us typically with several moduli, whose potential
is hard to explicitly compute and to minimize. Therefore it might be worthwhile
to derive modular flavor symmetries from less complex settings, such as
magnetized tori, where the background fluxes lead to chiral
fermions~\cite{Cremades:2004wa}. Such models seem to give rise to modular
flavor symmetries of the type discussed in this note~\cite{Kobayashi:2018bff}. These models are dual to $D$--brane
models~\cite{Cremades:2003qj}, and the couplings there can be mapped to
couplings on orbifolds~\cite{Abel:2003yx}.

All these arguments suggest that more efforts need to go into deriving the
modular flavor symmetries from string theory, or other higher--dimensional
models. It is only then one might control the K\"ahler potential well
enough to make controlled predictions.

As a side remark, let us also comment on the terminology.
In some of the recent literature, the transformation 
\begin{subequations}\label{eq:so-calledKahler}
\begin{align}
 \mathscr{W}(\Phi)&~\mapsto~\mathscr{W}(\Phi)\;,\\
 K(\Phi,\overline{\Phi})&~\mapsto~K(\Phi,\overline{\Phi})+f(\Phi)+\bar
 f(\overline{\Phi})\;,
\end{align}
\end{subequations}
where $f$ is a holomorphic function, is referred to as ``K\"ahler
transformation''. Since the Lagrangean of a supersymmetric theory is given by
\begin{align}
 \mathscr{L}~=~\int\!\D^4\theta\,K(\Phi,\overline{\Phi})
 +\left[\int\!\D^2\theta\,\mathscr{W}(\Phi)+\text{h.c.}\right]\;,
\end{align}
we note that it is invariant under \eqref{eq:so-calledKahler} just because
\begin{align}
 \int\!\D^4\theta\,f(\Phi)~=~\int\!\D^4\theta\,\bar f(\overline{\Phi})~=~0\;.
\end{align}
So \eqref{eq:so-calledKahler} is nothing but the statement that one can shift
the K\"ahler potential of a global supersymmetric theory by the real part of a
holomorphic function without changing a Lagrangean. This is not a K\"ahler
true transformation. K\"ahler transformations are formally written as~\cite{Wess:1992cp}
\begin{subequations}\label{eq:WessBaggerKahlerTransformations}
\begin{align}
 \mathscr{W}(\Phi)&~\mapsto~\mathrm{e}^{-f(\Phi)}\,\mathscr{W}(\Phi)\;,
 \label{eq:WessBaggerKahlerTransformationsW}\\
 K(\Phi,\overline{\Phi})&~\mapsto~K(\Phi,\overline{\Phi})+f(\Phi)+\bar
 f(\overline{\Phi})\;.
\end{align}
\end{subequations}
They have the virtue of leaving the scalar potential
\begin{equation}
 \mathscr{V}_\mathrm{SUGRA}~=~\mathrm{e}^K\,\left[
  K^{i\bar\jmath}\left(D_i\mathscr{W}\right)\,
  \left(\overline{D}_{\bar\jmath}\overline{\mathscr{W}}\right)
  -3|\mathscr{W}|^2
 \right]
\end{equation}
invariant. The K\"ahler
transformation~\eqref{eq:WessBaggerKahlerTransformations} does reduce
to~\eqref{eq:so-calledKahler} for dimensionful fields $\Phi$ at zeroth order in
$\Phi/\Lambda$ because of the suppression scale $\Lambda$ in the exponent
of $e^{-f}$. However, for dimensionless fields, such as $\Phi=T$ (or $\tau$)
(cf.~\cite[footnote~3]{Feruglio:2017spp}), no such suppression appears and
thus only~\eqref{eq:WessBaggerKahlerTransformations} is a proper K\"ahler 
transformation in this context. As mentioned above, it does not make
sense to expand in $T/\Lambda$, i.e.\ the point in field space at which $|T|$ is
small is not a point one may expand around. This observation becomes relevant
in constructions emerging from string theory, where the K\"ahler
transformations~\eqref{eq:WessBaggerKahlerTransformations}, and
not~\eqref{eq:so-calledKahler}, are symmetries of the theory.

\section{Summary}

Motivated by the striking observation that modular flavor symmetries allow one,
at some level, to successfully make several nontrivial predictions
\cite{Feruglio:2017spp,Criado:2018thu}, we have studied these models in some
more detail. We find that there are additional parameters which have not been
taken into account in the literature so far. The existence of these parameters
renders these models less predictive than previously thought.

Let us emphasize, though, that despite the existence of additional parameters,
the modular flavor symmetries continue to be highly interesting approach to the
flavor problem. It will be instrumental to derive them from a more complete
setting, in which one may hope to control the K\"ahler potential to a greater
degree.

\subsection*{Acknowledgments}

It is a pleasure to thank Patrick Vaudrevange for useful discussions. The work
of M.-C. C. was supported, in part, by the National Science Foundation, under
Grant No.\ PHY-1915005. The work of S.R.-S.\ was partly supported by DGAPA-PAPIIT grant IN100217, CONACyT
grants F-252167 and 278017, PIIF grant and the TUM August--Wilhelm Scheer
Program. The work of M.R.\ is supported by NSF Grant No.\ PHY-1719438.

\bibliography{Orbifold}
\bibliographystyle{NewArXiv}
\end{document}